\documentclass[final]{ias2}

\usepackage{graphicx} 
\usepackage{multirow}
\usepackage{array} 

\usepackage{hyperref} 

\begin{document}

\markboth{Searches for phenomena beyond the Standard Model at the LHC
with the ATLAS and CMS detectors}{Henri Bachacou}

\title{Searches for phenomena beyond the Standard Model at the LHC with the ATLAS and CMS detectors}

\author[sin]{Henri Bachacou on behalf of the ATLAS and CMS collaborations} 
\email{henri.bachacou@cea.fr}
\address[sin]{Irfu, CEA-Saclay, F-91191 Gif-sur-Yvette, France}

\begin{abstract}
The LHC has delivered several~fb$^{-1}$ of data in spring and summer 2011, opening
new windows of opportunity for discovering phenomena beyond the Standard Model.
A summary of the searches conducted by the ATLAS and CMS experiments based on about 1~fb$^{-1}$ 
of data is presented.
\end{abstract}


 
\maketitle


\section{Introduction}

The Standard Model (SM) has proven to be an incredibly successful theory over the past decades.
However successful, it is an effective theory that must break down above a certain 
energy scale, and there are strong theoretical arguments to believe that it breaks down
at the electroweak scale. For a theoretical review of this subject, see \cite{EWTheoryGautam} 
and the contribution in this conference from the same author.
In 2011, operating at a center-of-mass energy of 7~TeV in $pp$ collisions, the LHC has been able
to deliver several fb$^{-1}$ of data to both ATLAS~\cite{ATLAS} and CMS~\cite{CMS} detectors within a few months, 
allowing to extend the reach of searches for phenomena beyond the Standard Model well beyond the ones
carried by the TeVatron. 

This article presents some of the searches carried by ATLAS and CMS using up to 1.6~fb$^{-1}$ of data
on supersymmetry and exotic signatures. 
I will start with a summary of the searches for supersymmetry, followed by an overview of some
exotic searches, divided somewhat arbitrarily in three section: search for heavy resonances, 
search for strong gravity at the TeV-scale, and search for long-lived particles.
Unfortunately no deviation from the SM expectation is observed, but limits on many theories beyond 
the SM are improved significantly. 

Searches related to Higgs boson, top-antitop resonance and fourth generation quarks are described in 
other contributions of this conference~\cite{HiggsTalk, TopTalk, QCDTalk}.
Only a selection of results is shown here; all results can be found on the ATLAS~\cite{AtlasWeb}
and CMS~\cite{CMSWeb} web pages.

\section{Supersymmetry}

During the past decades, supersymmetry~\cite{SUSY1, SUSY2} has been considered the most promising extension of the SM.
The phenomenology of supersymmetry is very diverse, which requires a search strategy following
several classes of models and covering many signatures. 

In its most hoped for incarnation, supersymmetry is expected to be discovered at the LHC through 
pair production of supersymmetric particles decaying in a cascade of supersymmetric and SM particles.
If R-parity is conserved, the lightest supersymmetric particle (LSP) is stable and neutral, and 
the cascade ends with the production of LSP's, which escape the detector, producing
missing transverse momentum.

\begin{figure}[ht]
\begin{center}
\includegraphics[width=0.45\columnwidth]{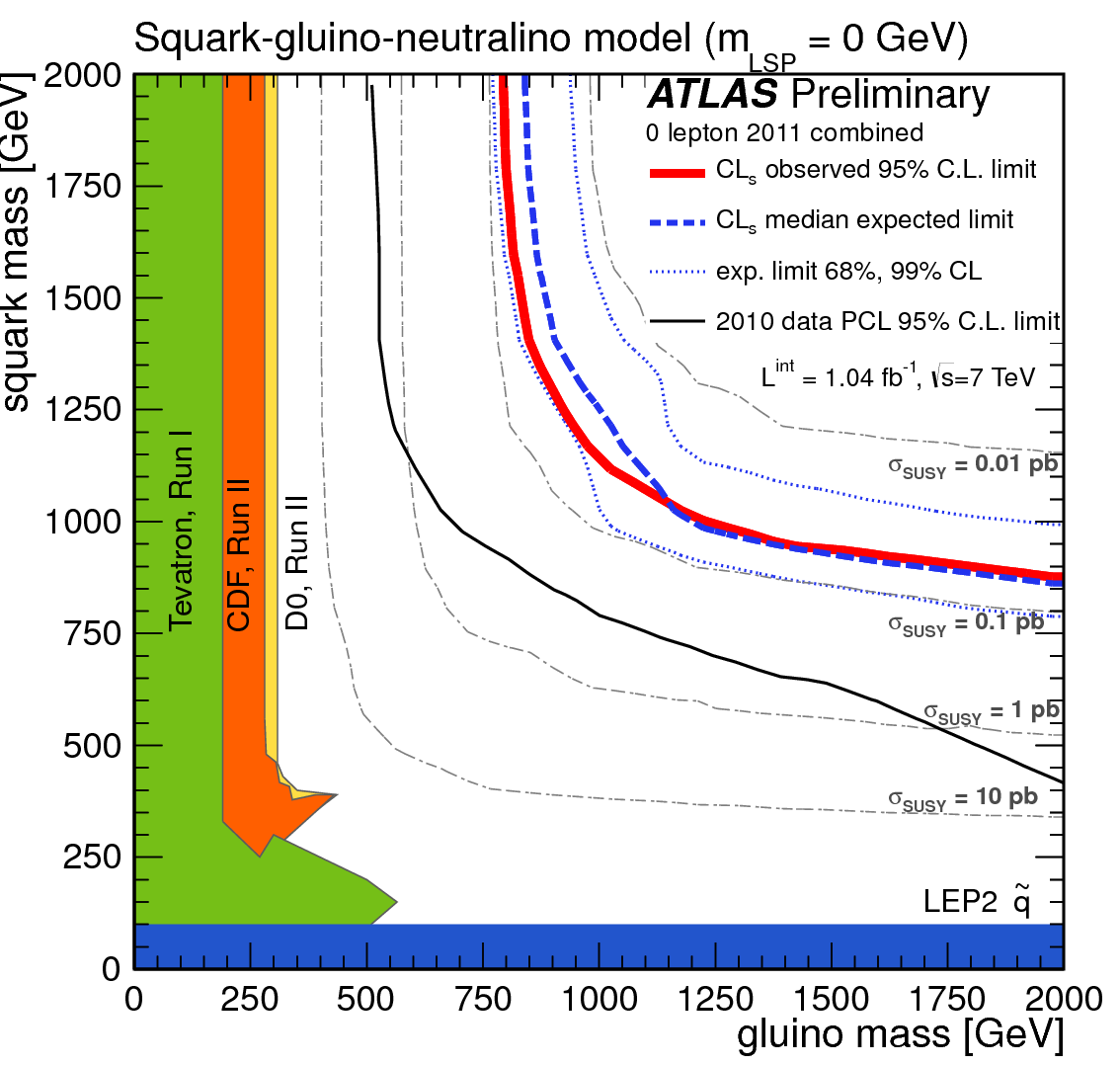}
\includegraphics[width=0.45\columnwidth]{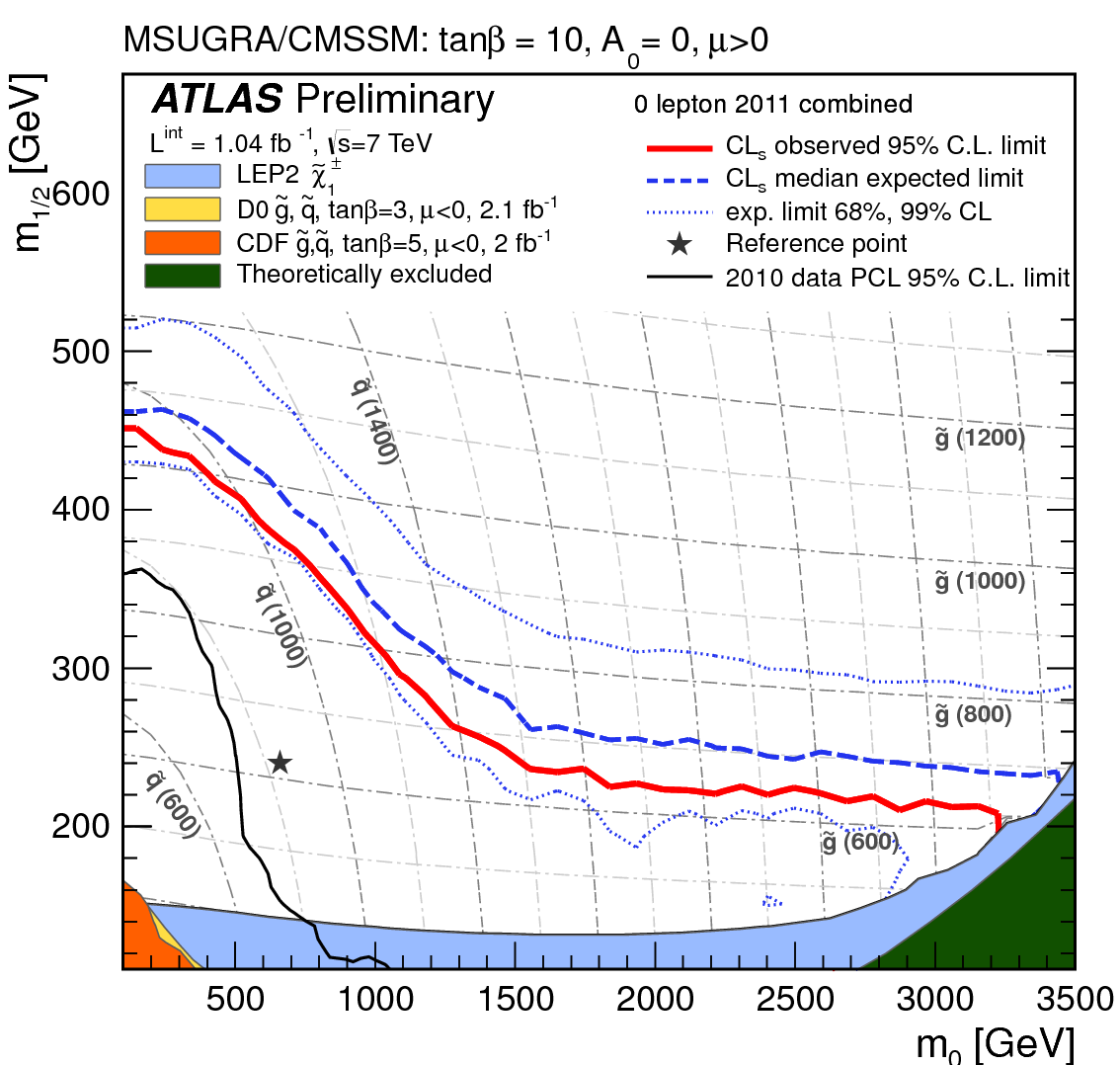}
\caption{Limits on supersymmetric models from the 0-lepton channel at ATLAS~\cite{ATLAS0lepton}.
Left: simplified model assuming only squark and gluino production, and a light LSP. 
Right: CMSSM/MSUGRA model.}
\label{fig:ATLASSUSYS0lepton}
\end{center}
\end{figure}

\begin{figure}[ht]
\begin{center}
\includegraphics[width=0.7\columnwidth]{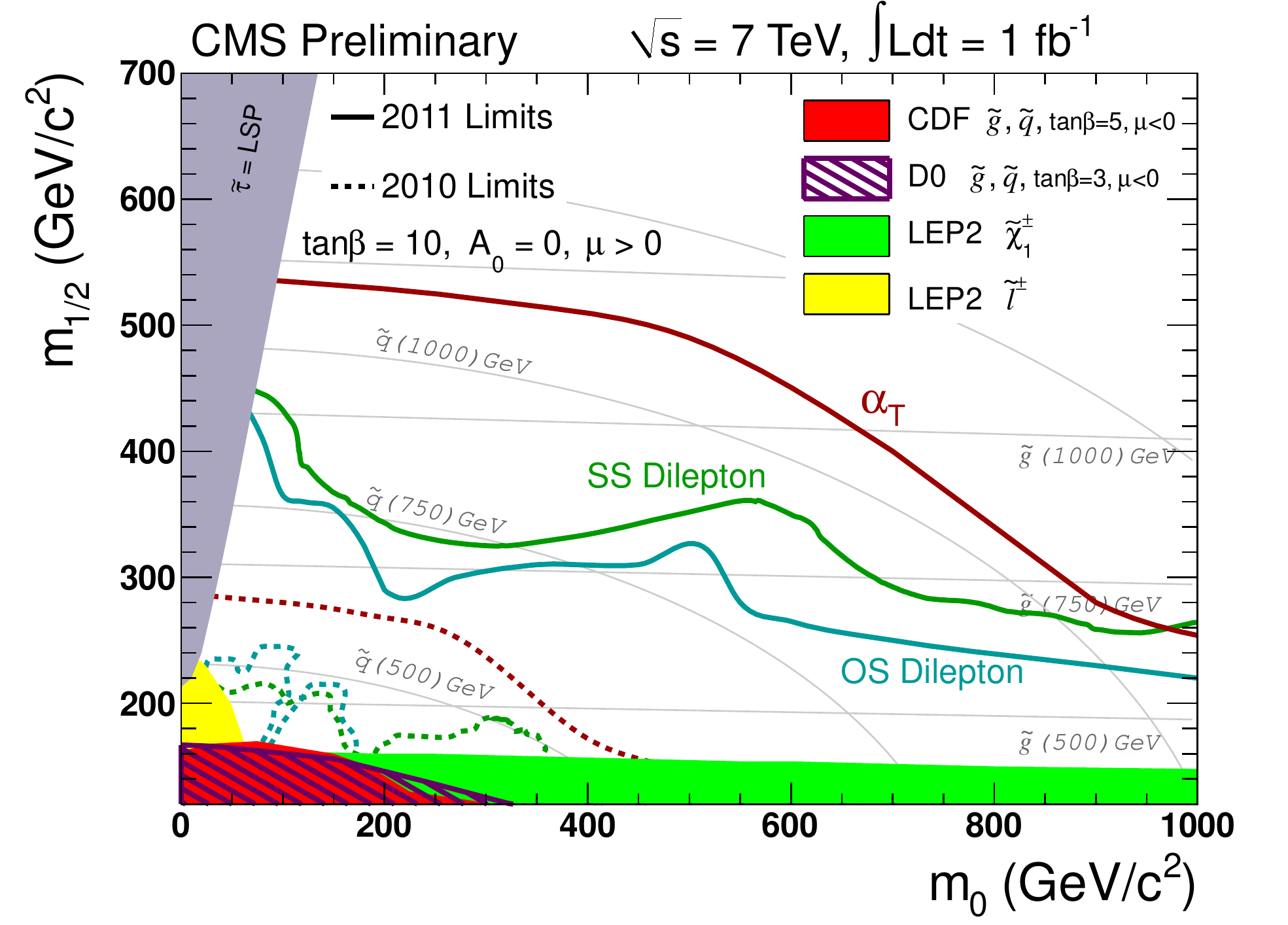}
\caption{Summary of the searches for supersymmetry at CMS on CMSSM/MSUGRA in the 0-lepton, 
1-lepton, and 2-lepton channel.}
\label{fig:CMSSUSYSummary}
\end{center}
\end{figure}

In $pp$ collisions, strongly coupled particles are much more likely to be produced, thus the production
of squarks and gluinos is expected to dominate, leading predominantly to a final state with
jets and missing transverse momentum. The ``workhorse'' of supersymmetry searches at the LHC is thus
the channel with large missing transverse momentum and jets of high transverse momentum.
No excess above the expected SM background is observed and limits are set on supersymmetric models. 
Figures~\ref{fig:ATLASSUSYS0lepton} and~\ref{fig:CMSSUSYSummary} show the limits from 
ATLAS~\cite{ATLAS0lepton} and CMS~\cite{CMS0lepton}.
In addition to setting limits on the CMSSM/MSUGRA model, ATLAS also presents a limit for a simplified
model assuming only squark and gluino production, and a cascade involving only quarks and gluons,
and the LSP. For equal masses of squarks and gluinos, a limit of about 1~TeV is set at 95\% CL. 

The cascade can also produce leptons through the decay of sleptons, charginos, or W/Z bosons.
Due to the smaller branching ratio, channels containing one~\cite{ATLAS1lepton, CMS1lepton} 
or more electron or muon 
are less sensitive to squark and gluino strong production, but are complementary 
to the fully hadronic channel, as shown in Figure~\ref{fig:CMSSUSYSummary} for the CMS results.
In the di-lepton channel, several strategies are employed: opposite-sign~\cite{CMSOSlepton} 
or same-sign~\cite{CMSSSlepton}, flavor subtraction~\cite{ATLAS2lepton}
to remove the flavor-correlated background, or explicit reconstruction of a Z produced in the 
cascade and decaying to a pair of muons or electrons~\cite{CMSZMET}.

\begin{figure}[ht]
\begin{center}
\includegraphics[width=0.7\columnwidth]{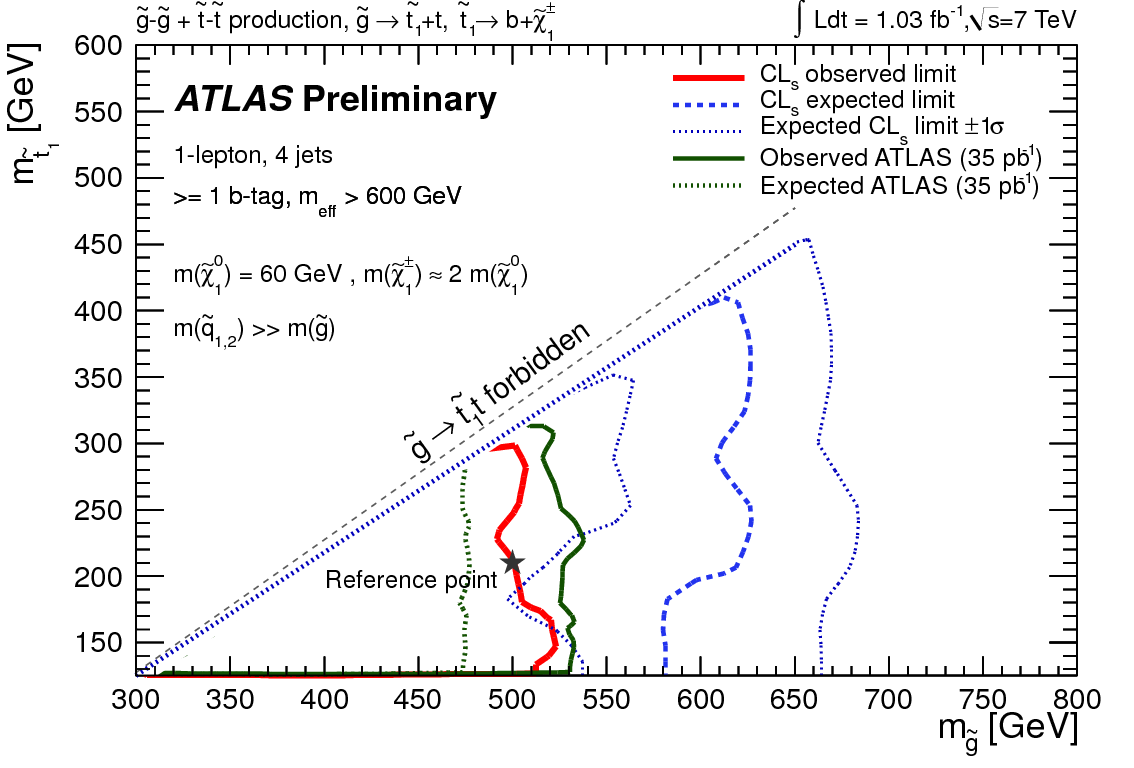}
\includegraphics[width=0.7\columnwidth]{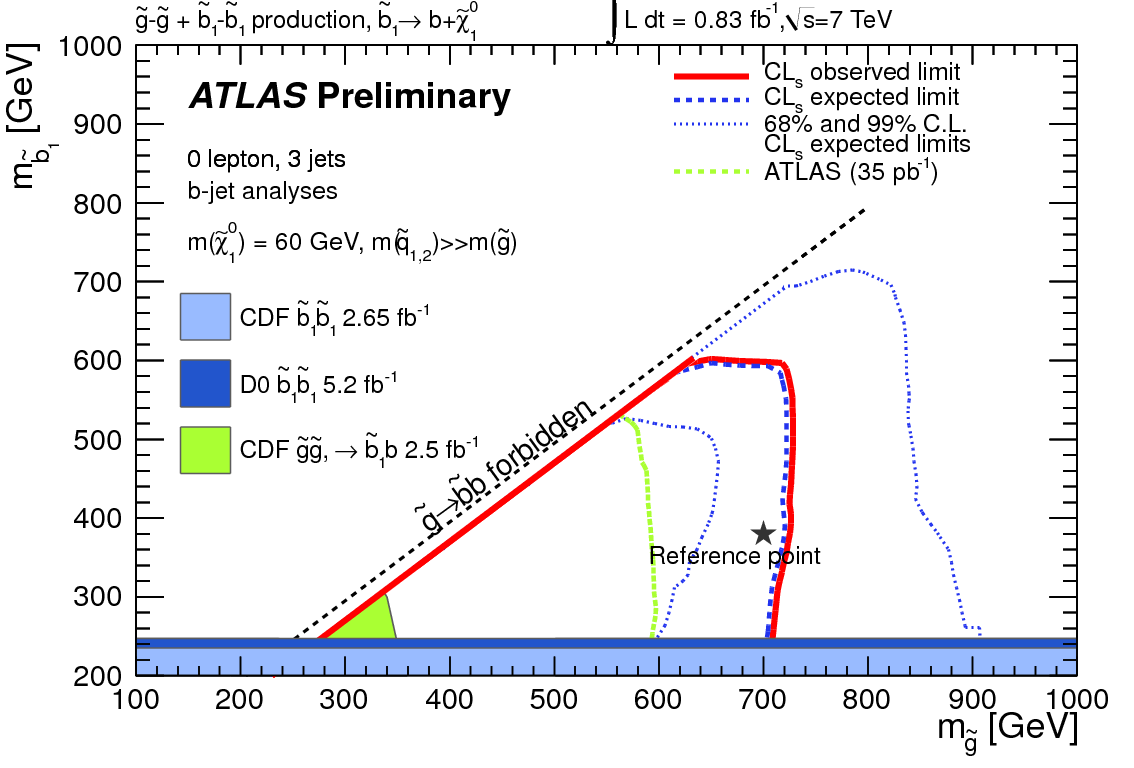}
\caption{Search for light third generation supersymmetric models.
Top: 1-lepton with at least one b-jet~\cite{ATLAS1lepton3G}.
Bottom: 0-lepton with b-jets~\cite{ATLAS0lepton3G}.}
\label{fig:ATLASSUSY3G}
\end{center}
\end{figure}

Of particular interest are scenarios in which the third generation of supersymmetric particles is 
much lighter than the others. The current luminosity allows to test such scenarios only through
production of gluinos decaying to stop or sbottom, leading to a final state of top and/or bottom quarks.
Assuming that the stop is the only light squark, 
gluino pair production leads to a complex final state containing top and bottom quarks.
Figure~\ref{fig:ATLASSUSY3G} (top) shows that this scenario is excluded for gluino masses
up to 500~GeV in the channel with one lepton and at least four jets, one of which identified as 
a b-jet~\cite{ATLAS1lepton3G}. 
Alternatively, if the only light squark is a sbottom, gluino pair production leads
to a final state with four b-jets and two LPS's; in this case, in the channel with at least 3 jets,
at least two of which identified as b-jets, gluino masses
are excluded up to 700~GeV, as shown on Figure~\ref{fig:ATLASSUSY3G} (bottom)~\cite{ATLAS0lepton3G}.
Additional luminosity will allow to search for direct production of third generation quarks and
gauginos.

\begin{figure}[ht]
\begin{center}
\includegraphics[width=0.53\columnwidth]{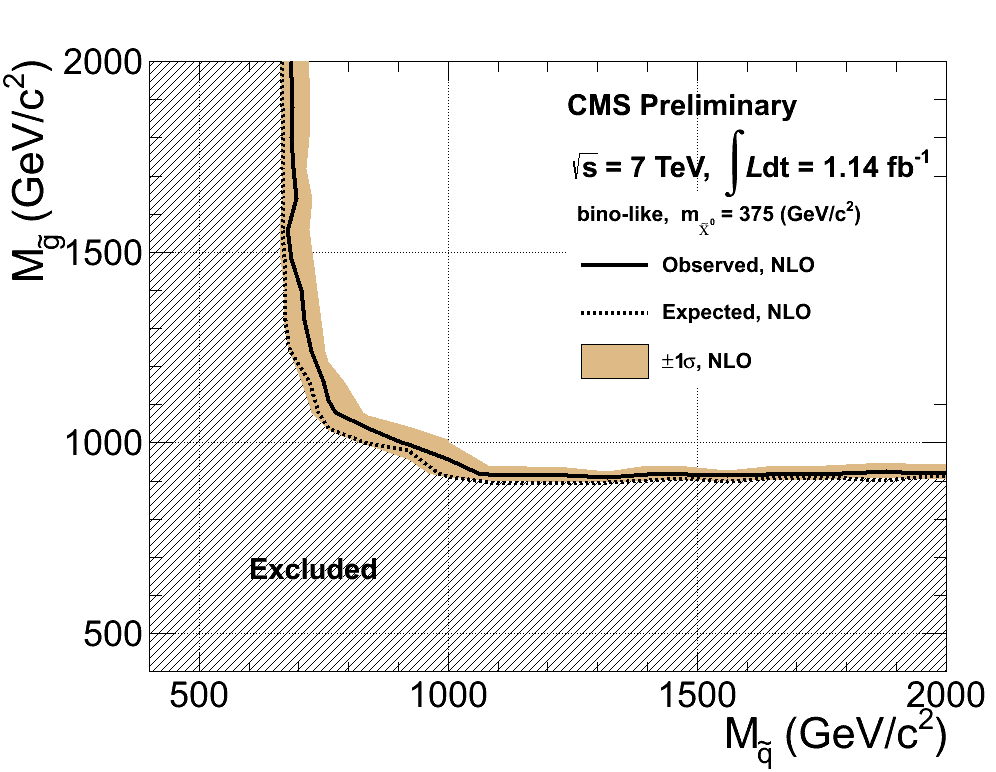}
\includegraphics[width=0.45\columnwidth]{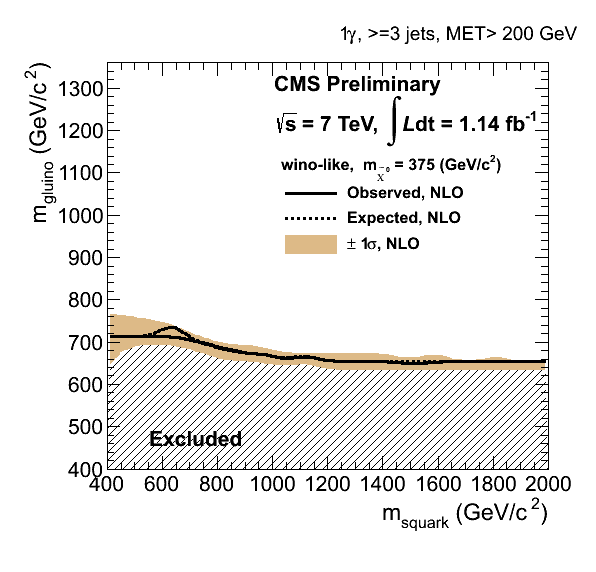}
\caption{Search for GMSB in the single-photon plus missing transverse momentum and diphoton 
 plus missing transverse momentum final states~\cite{CMSDiphotonMET}. Left: di-photon, bino-like interpretation.
Right: single photon, wino-like interpretation.}
\label{fig:CMSGMSB}
\end{center}
\end{figure}

In gauge-mediated supersymmetry breaking (GMSB) models~\cite{GMSB}, the LSP is the gravitino and the next lightest
supersymmetric particle (NLSP) is a neutralino or a chargino. This leads to a cascade ending with
photons and missing transverse momentum in the final state. 
CMS has looked for both single-photon and di-photon final states~\cite{CMSDiphotonMET}.
Results are shown in Figure~\ref{fig:CMSGMSB}. In the di-photon channel,
the result is also interpreted for the scenario of wino-like NLSP (neutralino
and chargino nearly degenerate in mass).
Universal Extra-Dimensions (UED) models~\cite{UED} predict cascades that are very similar to supersymmetry,
which allows to interpret the same analysis in both models~\cite{ATLASDiphotonMET}.

Supersymmetric signatures involving long-lived particles are discussed in the last section.


\section{Heavy resonances}

Heavy resonances are predicted by many extensions of the SM.
Some Grand Unified Theories~\cite{GUTE6} predict the existence of 
additional gauge bosons while Randall-Sundrum models with warped 
extra-dimensions~\cite{RS1, RS2} predict Kaluza-Klein excitations of the graviton.
Both lead to a narrow resonance decaying to a pair of fermions or bosons
with branching ratios varying widely depending on the model considered.

\begin{figure}[ht]
\begin{center}
\includegraphics[width=0.4\columnwidth]{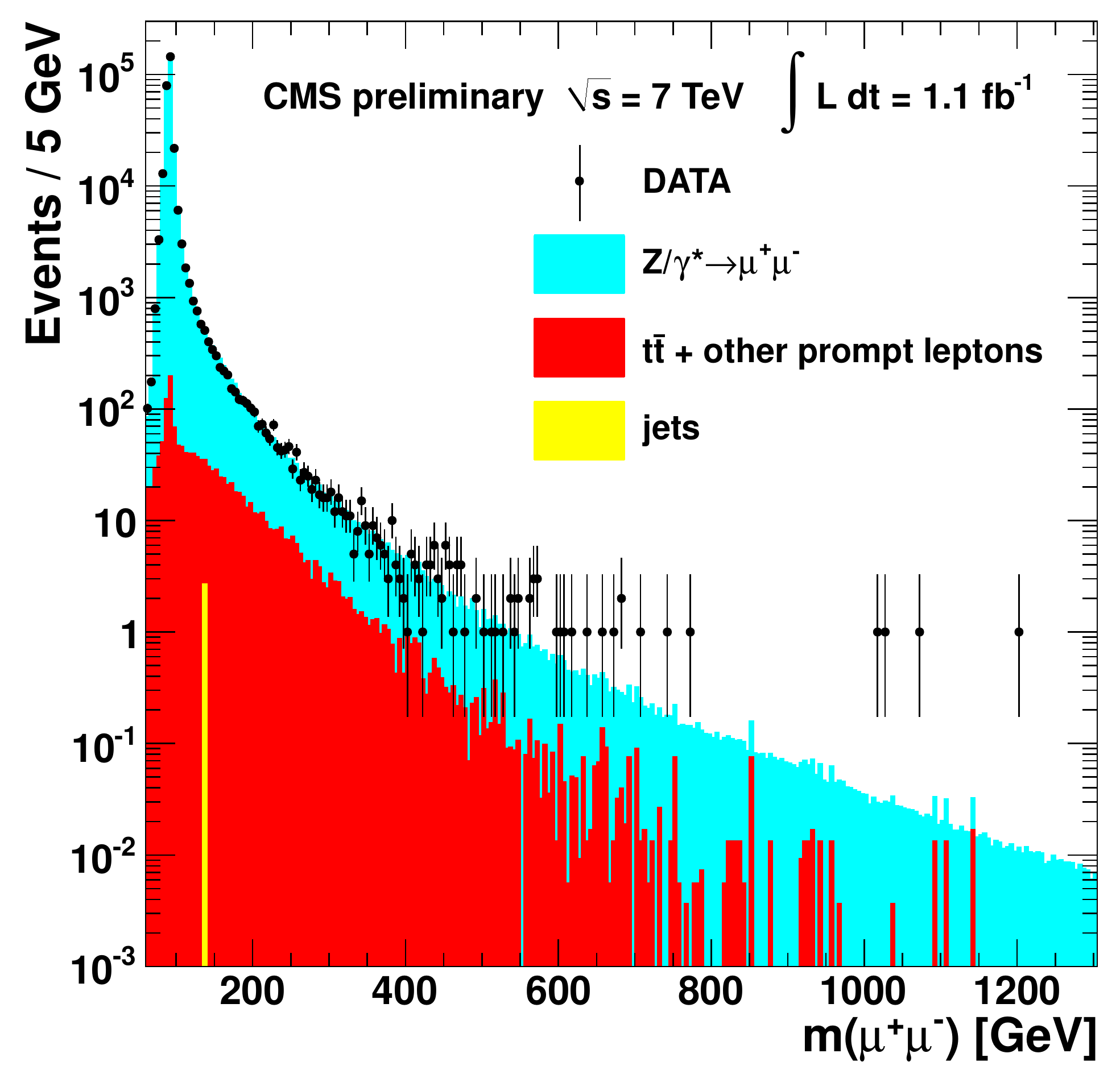}
\includegraphics[width=0.55\columnwidth]{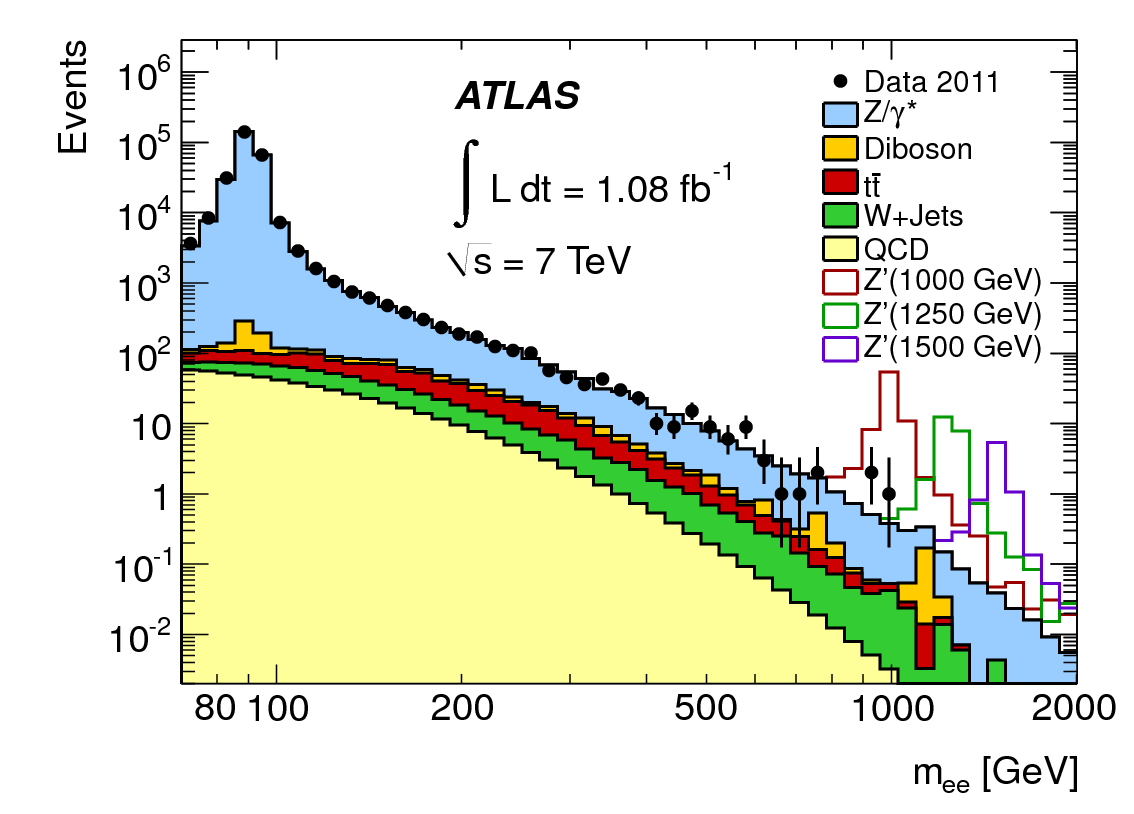}
\caption{Search for heavy resonances in the di-lepton channel. 
Left: reconstructed di-muon mass spectrum (CMS)~\cite{CMSZp}.
Right: reconstructed di-electron mass spectrum (ATLAS)~\cite{ATLASZp}.}
\label{fig:zp}
\end{center}
\end{figure}

\begin{figure}[ht]
\begin{center}
\includegraphics[width=0.4\columnwidth]{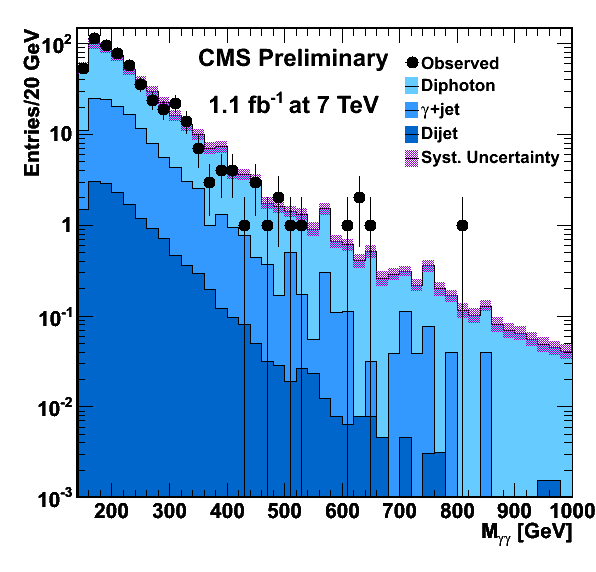}
\includegraphics[width=0.54\columnwidth]{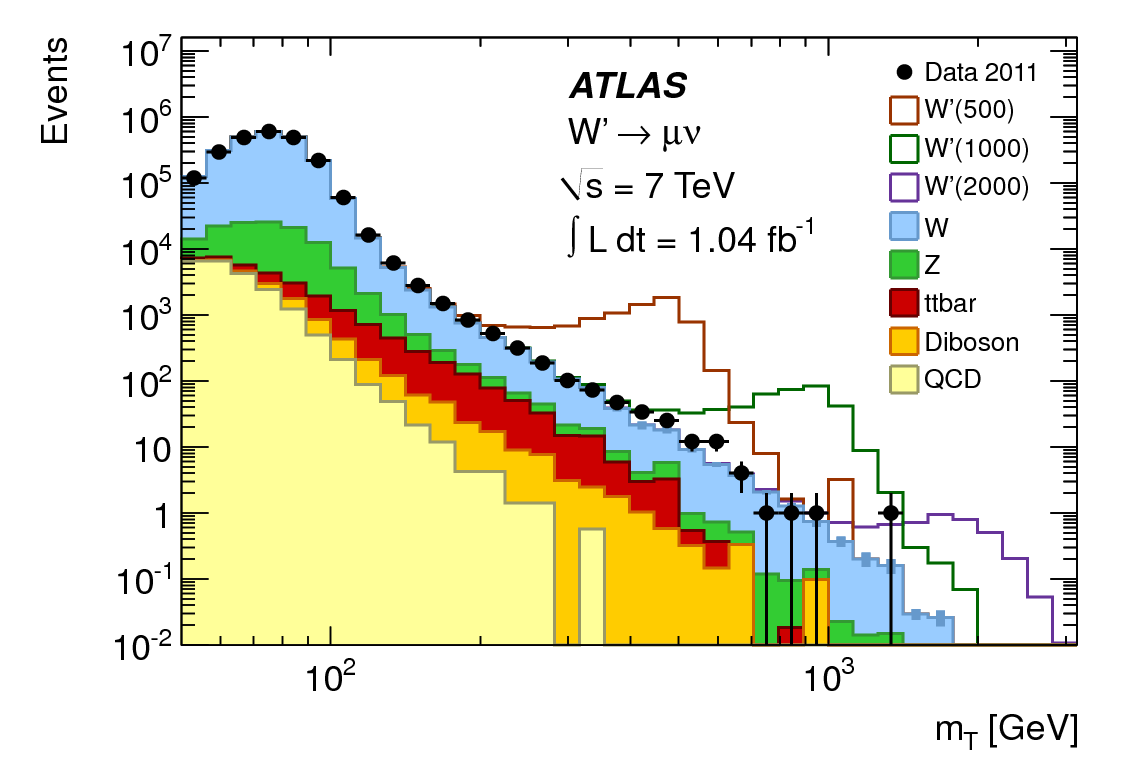}
\caption{
Left: Di-photon reconstructed mass spectrum (CMS)~\cite{CMSDiphoton}.
Right: reconstructed transverse mass in events with one muon and missing transverse momentum (ATLAS)~\cite{ATLASWp}.}
\label{fig:diph-wp}
\end{center}
\end{figure}

In the di-lepton channel (di-electron or di-muon)~\cite{ATLASZp, CMSZp}, 
a neutral gauge boson with the same couplings as the SM $Z^0$ (Sequential Standard Model Z'~\cite{SSM})
is excluded up to a mass of 1.9~TeV at 95\% CL. A Randall-Sundrum Kaluza-Klein graviton
with a coupling of $k/M_{Pl} = 0.1$ is excluded up to 1.8~TeV at 95\% CL combining the di-electron
and the di-muon channel,
and up to 1.7~TeV in the diphoton channel alone~\cite{CMSDiphoton}.
Figure~\ref{fig:zp} and figure~\ref{fig:diph-wp} (left) show the di-leptons and the di-photon mass spectra,
respectively.

A charged gauge boson ($W'$) is searched for in the $e\nu$ and $\mu\nu$ channels by reconstructing the transverse mass of
the lepton transverse momentum and the event missing transverse momentum.
Figure~\ref{fig:diph-wp} (right) shows the ATLAS transverse mass in $\mu\nu$ events.
A $W'$ with the same couplings as the SM $W$ (Sequential Standard Model W') is 
 excluded up to a mass of 2.3~TeV at 95\% CL when combining $e\nu$ and $\mu\nu$ 
channels~\cite{ATLASWp, CMSWp}. 
A W' is also expected to decay to $WZ$, which is also a channel of interest for 
Technicolor~\cite{TC} searches; CMS has looked for a narrow resonance in the final state
$WZ \rightarrow l\nu ll$ and excludes an SSM W' up a mass of 784~GeV and a techni-rho
up to a mass of 436~GeV in the parameter space $m_{\rho_{TC}} < m_{\pi_{TC}} + m_W$~\cite{CMSWpWZ}.

\begin{figure}[ht]
\begin{center}
\includegraphics[width=0.4\columnwidth]{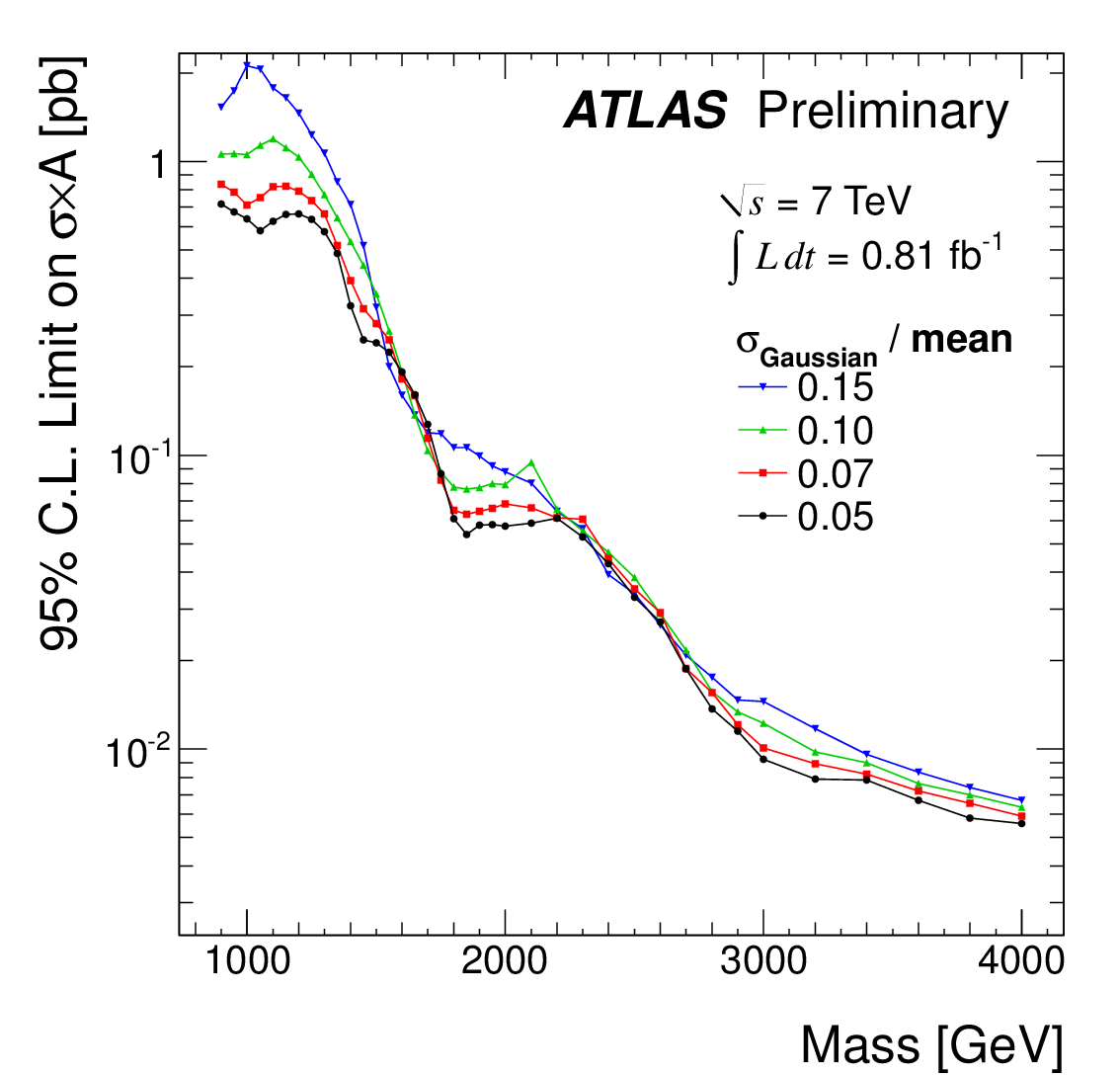}
\includegraphics[width=0.4\columnwidth]{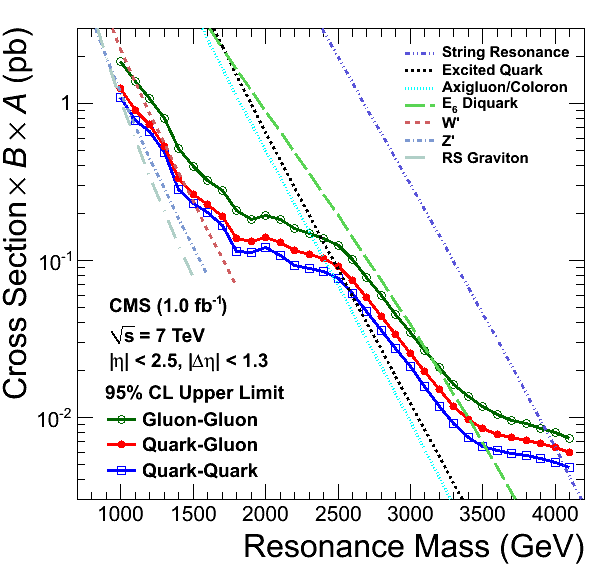}
\caption{95\% CL upper limits on the production cross-section times acceptance of resonances decaying to 
a pair of jets. Left: the limit is presented in a model-independent way as a function of
the full width (both physical and experimental) of the resonance (ATLAS)~\cite{ATLASdijet}. 
Right: limits on narrow resonances of type gluon-gluon, gluon-quark, and quark-quark are compared to 
various theoretical predictions (CMS)~\cite{CMSdijet}.}
\label{fig:dijet}
\end{center}
\end{figure}

A narrow resonance decaying to a pair of jets is also predicted by numerous models.
Considering the excited quark model ($q^*$)~\cite{qstar} as a benchmark, no narrow resonance 
in the di-jet system is observed up to 2.9~TeV at 95\% CL~\cite{ATLASdijet, CMSdijet}.
Figure~\ref{fig:dijet} shows the ATLAS model-independent limit on the cross-section for several hypotheses
of the resonance width, and the CMS limits on several models depending on the nature of the jet (quark jet or gluon jet).

\begin{figure}[ht]
\begin{center}
\includegraphics[width=0.57\columnwidth]{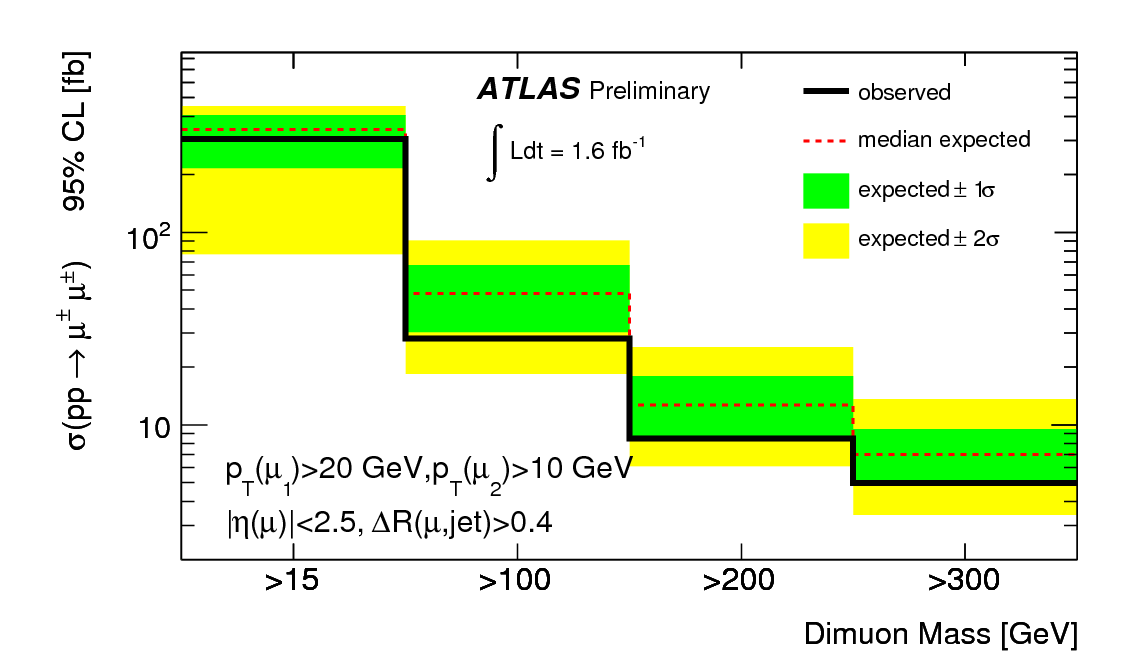}
\includegraphics[width=0.42\columnwidth]{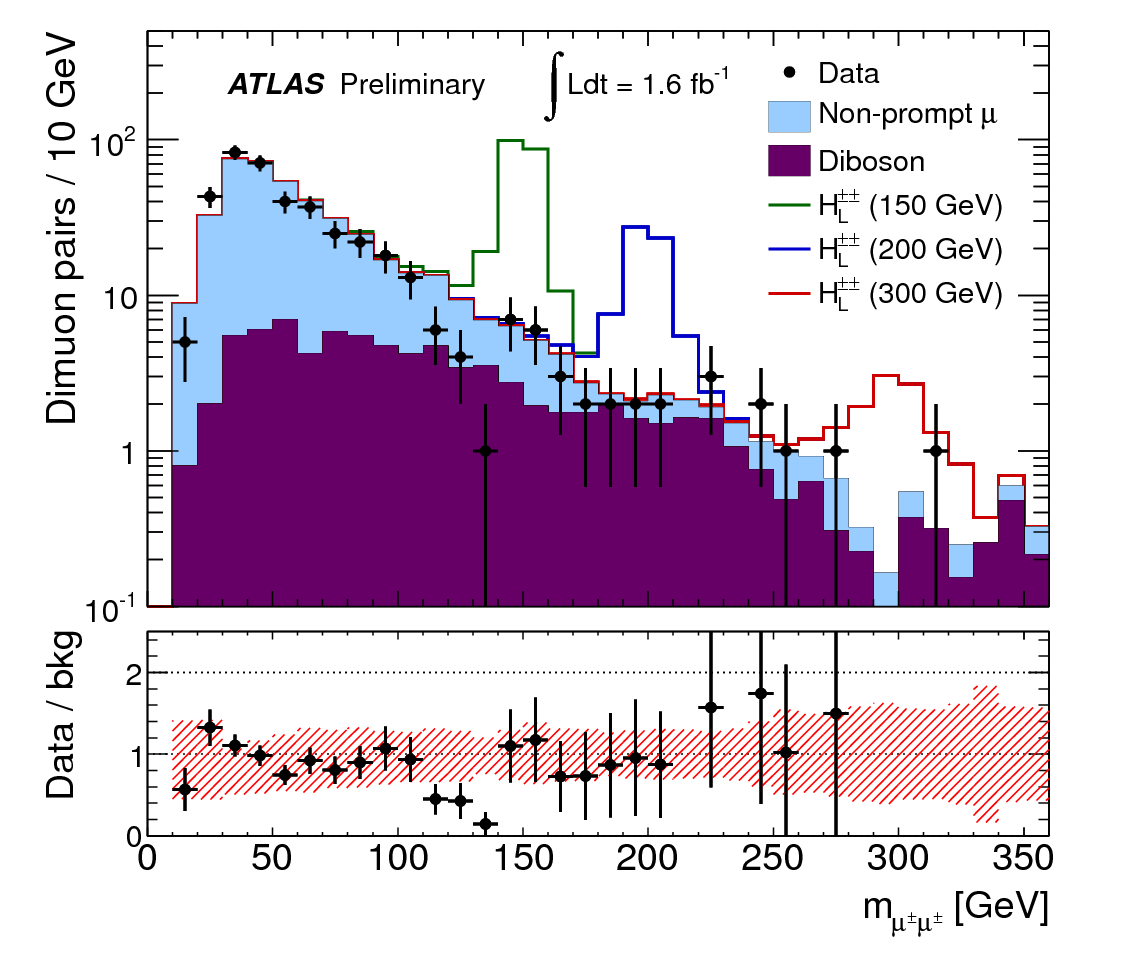}
\caption{Left: model-independent fiducial cross section
limit on the production of same-sign di-muon pairs~\cite{ATLASinclSS}.
Right: reconstructed mass of same-sign muon pairs in ATLAS and expected
doubly-charged Higgs signal of various masses~\cite{ATLASHpp}.}
\label{fig:ATLASSameSign}
\end{center}
\end{figure}

A heavy particle decaying to a pair of charged leptons of same-sign, such as a doubly-charged Higgs,
would be a striking signature of physics beyond the SM. More generally, final states including a pair of charged leptons of same-sign are predicted
by many BSM models (including supersymmetry, same-sign top production, fourth generation b', heavy
Majorana neutrino, etc...) and enjoy a very small SM background. 
Thus an inclusive search for same-sign di-lepton pair is very sensitive to a wide range of models, 
and thanks to the small background is almost as sensitive as a search optimized for a particular model. 
With 1.6~fb$^{-1}$ of integrated luminosity, ATLAS sets a 
model-independent limit on the fiducial cross-section of isolated pairs of same-sign muons as
a function of the di-lepton pair mass~\cite{ATLASinclSS} as shown on Figure~\ref{fig:ATLASSameSign}. 
The same mass spectrum is used to search for a narrow resonance,
allowing to exclude doubly-charged Higgs pair production up to a mass of 375~GeV in the 
left-handed coupling triplet model~\cite{ATLASHpp}.

\section{Strong gravity}

Theories of extra-dimension are a possible answer to the hierarchy problem.
In the large extra-dimension ADD model~\cite{ADD}, gravity is allowed to propagate into extra-dimensions,
thus appearing weak at (spatial) scales much larger than the scale of the extra-dimensions,
but possibly becoming strong at a scale of 1/TeV. The fundamental mass scale $M_D$ 
at which gravity becomes strong is related to the Planck scale via
$m^2_{Pl} = m^{2+n}_D R^n$ where $n$ is the number of extra-dimensions, and $R$ is the size of the
extra-dimension, and can indeed be close to the TeV scale for well-chosen values of $n$ and $R$.

\begin{figure}[ht]
\begin{center}
\includegraphics[width=0.56\columnwidth]{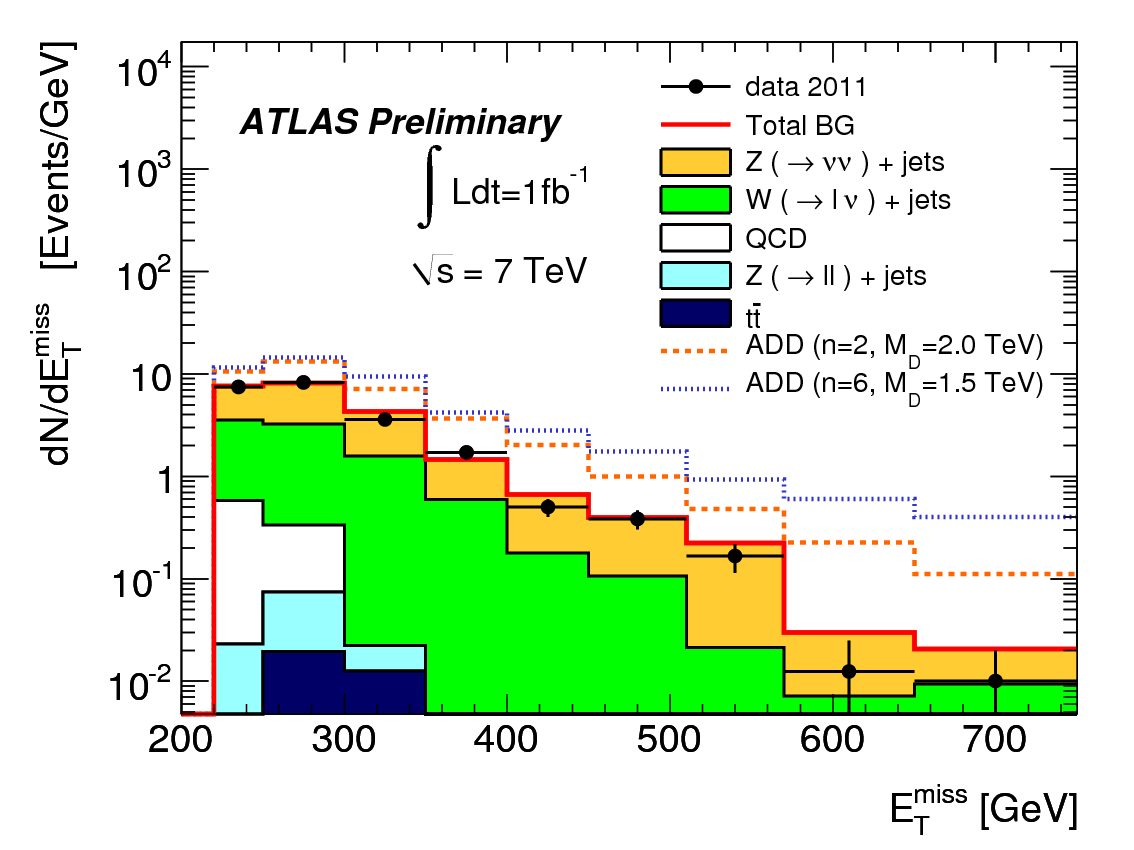}
\includegraphics[width=0.43\columnwidth]{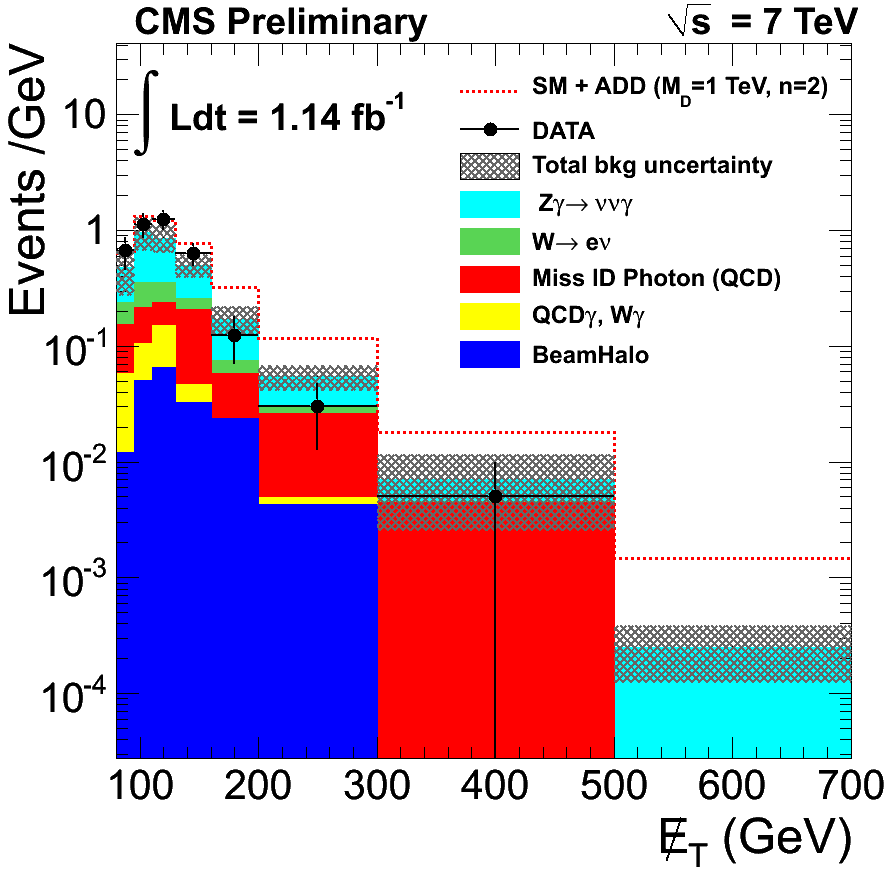}
\caption{Missing transverse momentum in ATLAS mono-jet (left) and CMS mono-photon (right) analyses~\cite{ATLASmonojet, CMSmonojet}.}
\label{fig:mono}
\end{center}
\end{figure}

A promising signature at colliders is the production of a single graviton
escaping the detector and recoiling against a jet or a photon, leading
to mono-jet~\cite{ATLASmonojet, CMSmonojet} or mono-photon~\cite{CMSmonophoton} 
final states with large missing transverse momentum.
Figure~\ref{fig:mono} shows the missing transverse momentum spectrum in the ATLAS mono-jet (left)
and CMS mono-photon (right) analyses.
Thanks to a larger cross-section the mono-jet channel gives the most stringent limits,
excluding $M_D$ up to 3.7~TeV for $n=2$ and 2.3~TeV $n=6$ (conservatively assuming LO cross-sections).

Another signature of ADD extra-dimensions is a non-resonant enhancement of expected di-lepton and di-photon events 
at high invariant mass through virtual graviton exchange. CMS has searched for deviations in the 
di-muon~\cite{CMSADDdimuon} and di-photon~\cite{CMSADDdiphoton} spectra, with a sensitivity similar to the monojet channel.

\begin{figure}[ht]
\begin{center}
\includegraphics[width=0.45\columnwidth]{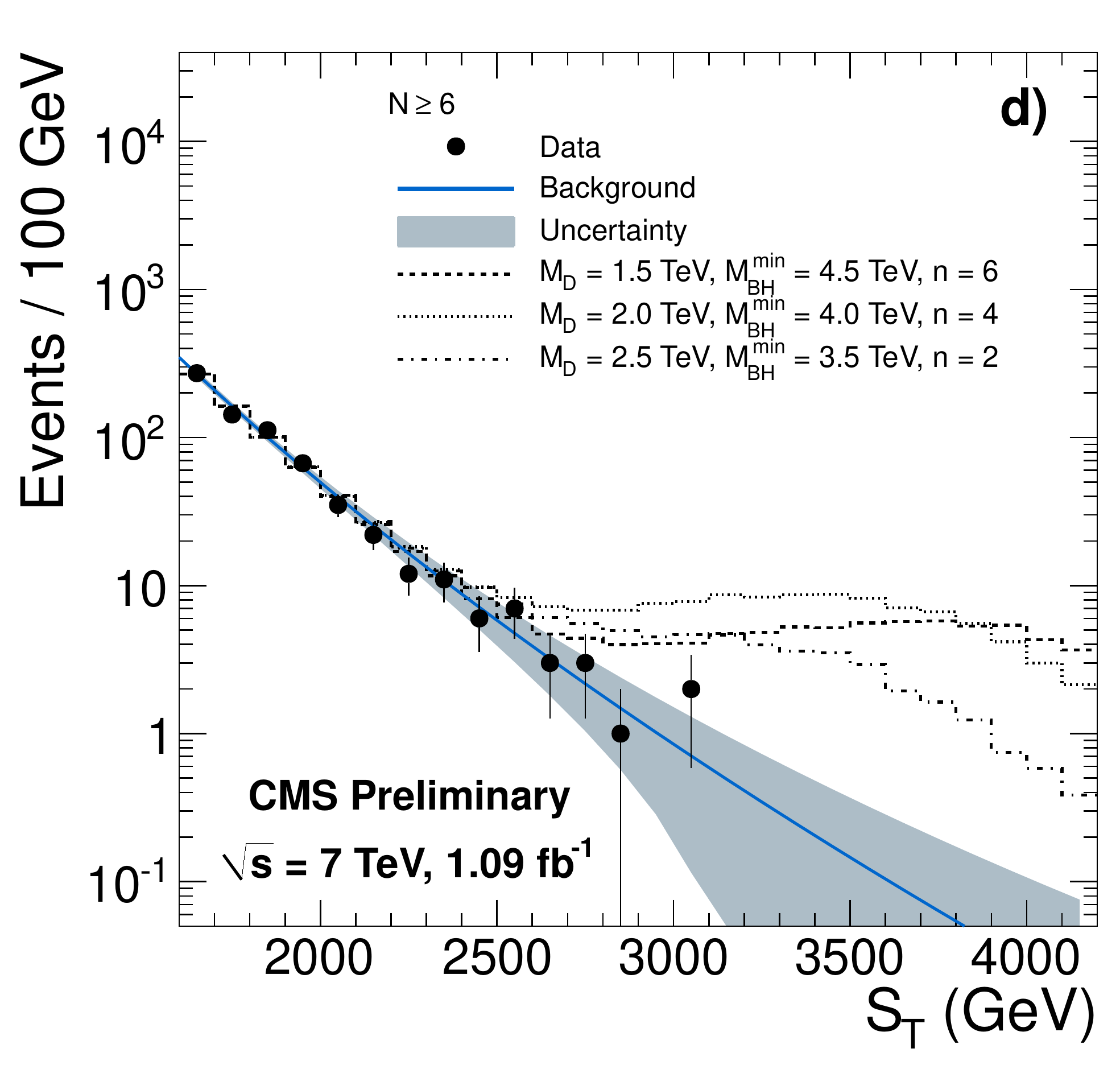}
\includegraphics[width=0.45\columnwidth]{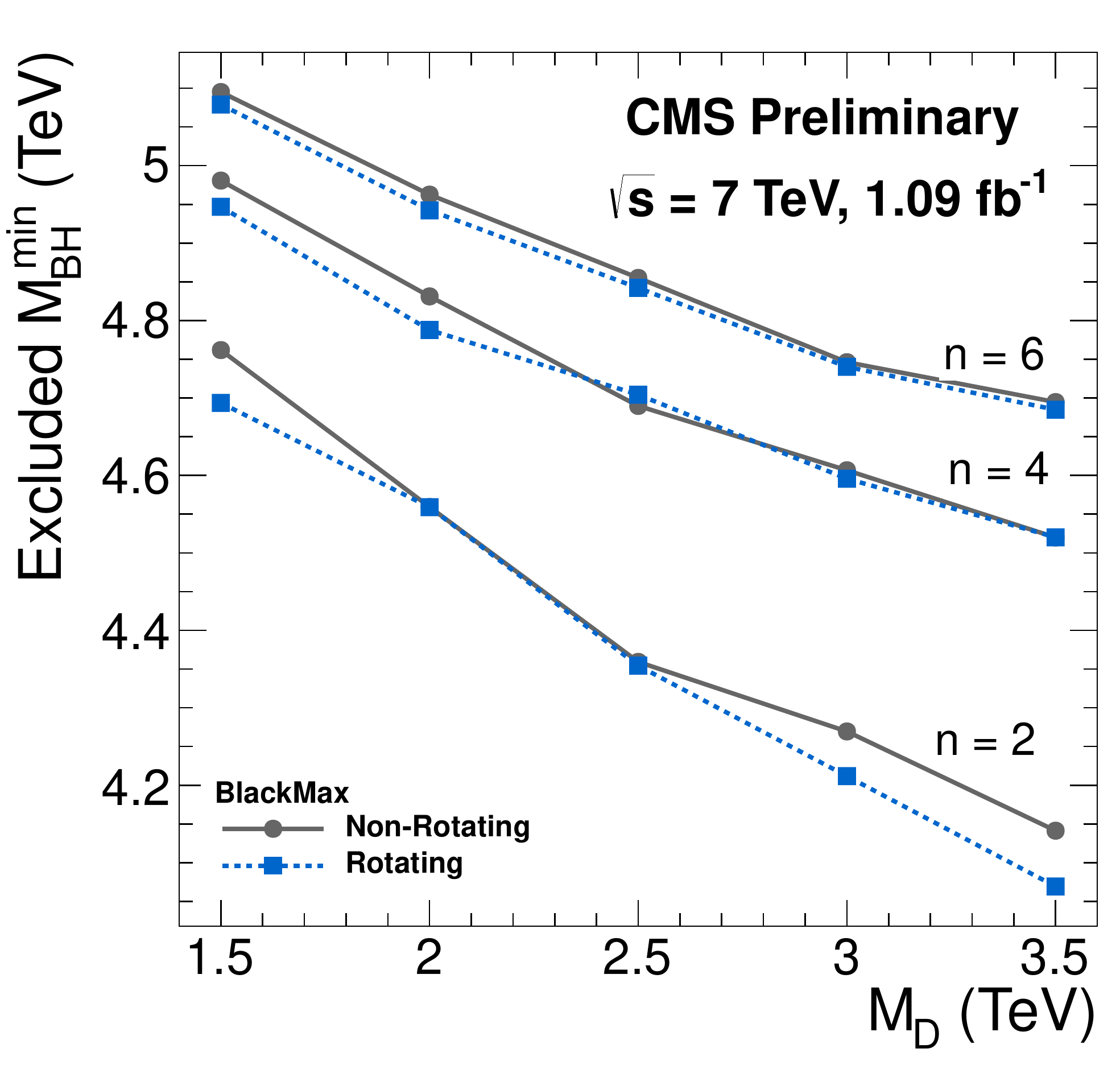}
\caption{Multi-object search for microscopic black-holes at CMS~\cite{CMSblackhole}.
Left: Scalar sum of all objects transverse momentum in events with at least six objects.
Right: Limit on black-hole mass as a function of $M_D$, number of extra-dimensions,
for two black-hole models.}
\label{fig:CMSbh}
\end{center}
\end{figure}

Finally, if gravity becomes strong at the TeV scale, microscopic black-holes may be produced
at the LHC. Due to our lack of understanding of quantum gravity, it is impossible to make precise predictions
of such phenomena. However one can expect such objects to decay democratically and isotropically, 
leading to a final state with a large multiplicity of high-momentum particles, and a high content of leptons.
Several channels have been considered: multi-jet~\cite{ATLASblackholeMJ}, 
same-sign di-muon with a high track multiplicity~\cite{ATLASblackholeSS}, 
and multi-object~\cite{CMSblackhole} (where an object refers to an electron, a muon, a photon, or a jet, and
a large number of objects is required in the event). 
In the latter case, CMS sets limits on black-hole masses up to 4-5~TeV for some classes of models.
Figure~\ref{fig:CMSbh} shows the $S_T$ variable, defined as the scalar sum of the transverse momentum of all objects
in the event, for events with at least six objects (left), and the limits achieved on the black-hole mass (right).

\section{Long-lived particles}

Several extensions of the SM, including Hidden Valley models~\cite{HV}, and supersymmetry in some scenarios~\cite{split}, 
predict the existence of long-lived heavy particles.
In the case of supersymmetry, a long-lived gluino or squark hadronizes into hadronic states called R-hadrons.
The experimental signature depends strongly on the property of the particle, and in particular its life-time.
If the life-time is short (between 1 ps and several ns) the particle decays within the detector in time with the 
collision that produced it; in this case it is possible to identify the decay thanks to dedicated vertexing~\cite{ATLASdisplaced}.

\begin{figure}[ht]
\begin{center}
\includegraphics[width=0.7\columnwidth]{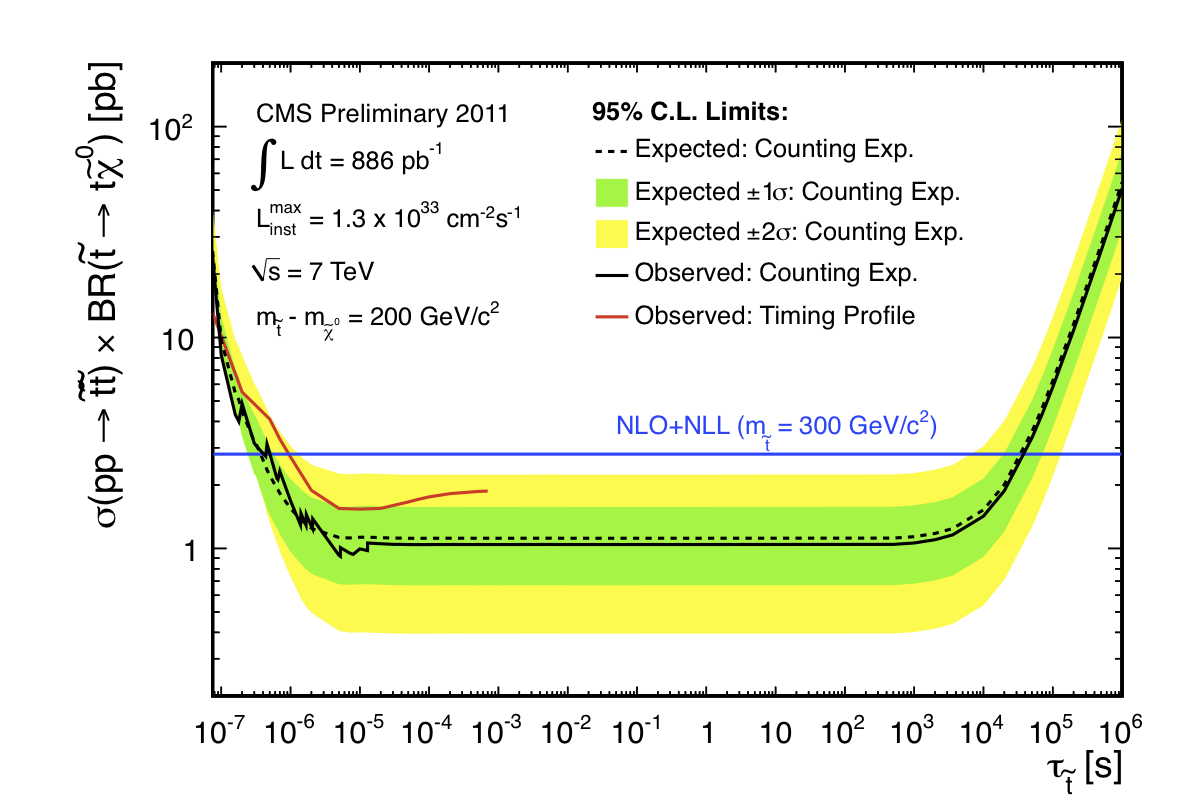}
\caption{Search for out-of-time decay of heavy long-lived particles stopped in the detector:
limit on the cross section times branching ratio as a function of the stop life-time~\cite{CMSoutoftime}.}
\label{fig:CMSLongLived}
\end{center}
\end{figure}

If the particle life-time is much longer than 1~ns there is no hope to see it decay in the detector. If the particle
is charged, it is possible to take advantage of the properties of a slow heavy particle and identify it thanks to
high energy loss in the tracking detectors and long time-of-flight~\cite{CMSslow}.
Alternatively, for a life-time up to about 1 month, if the particle is stopped within the detector, it is possible
to observe its decay long after the collision that produced it occurred~\cite{CMSoutoftime}.
Figure~\ref{fig:CMSLongLived} shows the limit on the production of long-lived scalar top stopping in the detector
and decaying out-of-time; the analysis is sensitive over 13 orders of magnitudes, from 100 ns to 1 month.

\section{Conclusion}

The LHC has performed exceptionally well and has provided ATLAS and CMS with more luminosity than 
expected. Many searches for physics beyond the SM have been conducted with up to 1.6~fb$^{-1}$, covering
a wide range of signatures. Unfortunately no deviation from the SM has been observed so far.
Supersymmetry in its most hoped-for incarnation is starting to be pushed
to the boarder of fine-tuning: in the framework of the CMSSM, supersymmetry is excluded up to a mass of 1~TeV in the
(optimistic) scenario of equal squark-gluino mass. This opens the field to variations of supersymmetry that require
more luminosity or new search strategies. Heavy gauge bosons are excluded up to masses of about 2~TeV, while
quark compositeness is tested up to 3~TeV.

\acknowledgments
I want to acknowledge the immense work of the LHC, ATLAS, and CMS collaborations that was 
required to achieve the results presented here.
I am very grateful to the organizers for a very fruitful conference and for an unforgettable stay in Mumbai.

\bibliographystyle{pramana}
\bibliography{references}

\end{document}